\documentclass[aps,10pt,showpacs,twocolumn]{revtex4}
\usepackage{amsmath, amssymb, lipsum}
\usepackage{graphics, graphicx, epsfig, color, subfigure}
\begin{document}

\newcommand{\ket}[1]{| #1 \rangle }
\newcommand{\bra}[1]{\langle #1 | }
%
%

\title{Collapses and Avoiding Wave Function Spreading}


\author{Aharon Casher}
\affiliation{Institute for Quantum Studies, Chapman University, Orange, CA USA \\and\\ School of Physics and Astronomy, Tel Aviv University, Tel Aviv, Israel}

\author{Shmuel Nussinov}
\email{nussinov@gmail.com}
\affiliation{Institute for Quantum Studies, Chapman University, Orange, CA USA \\
		and\\
School of Physics and Astronomy, Tel Aviv University, Tel Aviv, Israel}

\author{Jeff Tollaksen}
\email{tollaksen@chapman.edu}
\affiliation{Institute for Quantum Studies, Chapman University, Orange, CA USA \\
and\\
		Schmid College of Science and Technology, Chapman University, Orange, CA USA}

\begin{abstract}
We address the impossibility of achieving exact time reversal in a system with many degrees of freedom. This is a particular example of the difficult task of ``aiming" an initial classical state so as to become a specific final state.
We also comment on the classical-to-quantum transition in any non-separable closed system of $n \geq 2$ degrees of freedom. Even if the system is initially in a well defined WKB, semi-classical state, quantum evolution and, in particular, multiple reflections at classical turning points make it completely quantum mechanical with each particle smeared almost uniformly over all the configuration space. The argument, which is presented in the context of $n$ hard discs, is quite general. Finally, we  briefly address more complex quantum systems with many degrees of freedom  and ask when can they provide an appropriate environment to the above simpler systems so that quantum spreading is avoided by continuously leaving ``imprints" in the environment.  We also discuss the possible connections with the pointer systems that are needed in the quantum-to-classical ``collapse" transitions.
\end{abstract}

\maketitle

\section{Introduction}

The ``collapse" of a wave function into one of the eigenstates $\ket{a_i}$ of the
operator $\hat{A}$ after a ``hard"  von Neumann~\cite{vn} measurement,  is an outstanding puzzle in the interpretation of quantum mechanics (where $\hat{A}\ket{a_i}=a_i\ket{a_i}$). The von Neumann measurement correlates components, $\ket{a_i}$, 
of the initial wave function, $\ket{\Psi}$, of the measured system  with specific states $\ket{\Phi_i}$ of the measuring device or ``pointer." 
 von Neumann taught us that in order to model the measurement of an observable $\hat{A}$, we use an interaction Hamiltonian $H_{\mathrm{int}}$ of the form 
$H_{\mathrm{int}}=-\lambda(t)\hat{Q}_{\mathrm{md}}\hat{A}$
where $\hat{Q}_{\mathrm{md}}$ is an observable of the measuring device (e.g. the position of the pointer) and $\lambda(t) $ is a coupling constant which determines the duration and strength  of the measurement.
For a ``hard" or impulsive  measurement we need the coupling to be strong and short and thus take $\lambda (t)\neq0$ only for
   $t\in(t_0-\varepsilon,t_0+\varepsilon)$ and set
   $\lambda=\int_{t_0-\varepsilon}^{t_0+\varepsilon}\lambda(t)dt$. We may then neglect the time evolution
   given by $H_{\mathrm{s}}$ and $H_{\mathrm{md}}$ in the complete Hamiltonian $H=H_{\mathrm{s}}+_{\mathrm{md}}+H_{\mathrm{int}}$.
Using the Heisenberg equations-of-motion for the momentum $\hat{P}_{\mathrm{md}}$ of the measuring device (conjugate to the position $\hat{Q}_{\mathrm{md}}$), we see that $\hat{P}_{\mathrm{md}}$ evolves according to $\frac{d\hat{P}_{\mathrm{md}}}{d t}=\lambda (t) \hat{A}$.
Integrating this, we see that $P_{\mathrm{md}}(T)-P_{\mathrm{md}}(0)=\lambda \hat{A}$, where $P_{\mathrm{md}}(0)$ characterizes the initial state of the measuring device and $P_{\mathrm{md}}(T)$ characterizes the final.  
To make a more precise determination of $\hat{A}$ requires that the shift in $P_{\mathrm{md}}$, i.e. $\delta P_{\mathrm{md}}=P_{\mathrm{md}}(T)-P_{\mathrm{md}}(0)$, be distinguishable from it's uncertainty, $\Delta P_{\mathrm{md}}$.  This occurs, e.g., if $P_{\mathrm{md}}(0)$ and $P_{\mathrm{md}}(T)$ are more precisely defined and/or if $\lambda$ is sufficiently large.  
However, under these conditions (e.g. if the measuring device approaches a delta function in $P_{\mathrm{md}}$),
 then the disturbance or back-reaction on the system is increased due to a larger  $H_{\mathrm{int}}$, the result of the larger $\Delta Q_{\mathrm{md}}$ ($\Delta Q_{\mathrm{md}}\geq\frac{1}{\Delta P_{\mathrm{md}}}$).
When $\hat{A}$ is 
measured in this way, then any operator $\hat{O}$ ($[\hat{A},\hat{O}]\neq 0$) is disturbed because it evolved according to $\frac{d}{dt}{\hat{O}}=i\lambda(t)[\hat{A},\hat{O}]\hat{Q}_{\mathrm{md}}$, and since $\lambda\Delta Q_{\mathrm{md}}$ is not zero, $\hat{O}$ changes in an uncertain way proportional to $\lambda\Delta Q_{\mathrm{md}}$. 

\bigskip

In the Schr\"odinger picture, the time evolution operator for the
   complete system from $t=t_0-\varepsilon$ to $t=t_0+\varepsilon$ is $\exp\{-i\int_{t_0-\varepsilon}^{t_0+\varepsilon}H(t)dt\}=\exp\{-i\lambda\hat{Q}_{\mathrm{md}}\hat{A}\}$.  This shifts $P_{\mathrm{md}}$.  If before the measurement the system was in a superposition of eigenstates of $\hat{A}$, i.e. $\ket{\Psi} = \Sigma_i c_i\ket{a_i}$, then  the initial product state, $\ket{\Psi}\ket{\Phi}$, of the quantum and pointer systems, is transformed during the measurement process into the entangled state:
\begin{equation}
  \Sigma_i c_i\ket{a_i} \Phi_i(P)\rangle .
\label{entangledstateEq1}
\end{equation}
 with $\Phi_i(P)$ the wave function of the measuring device, i.e. the pointer, which had been shifted, via the von Neumann  interaction by the measured eigenvalue $a_i$.  In other words, the measuring device will also be in a ``superposition" corresponding to the system. This leads to the famous ``quantum measurement problem," discussed in ~\cite{leggett}.  A conventional solution to this problem is to argue that because the measuring device is macroscopic, it cannot be in a
   superposition, and so it will ``collapse" into one of these states and the system will collapse
   with it.  Without getting into physical theories as to how this might happen, there is general agreement as to the mathematical description of the net effect of the collapse: it converts the wave function of eq. (\ref{entangledstateEq1}) into a density matrix:
\begin{equation}
 \rho = \Sigma{|c_i|^2} \ket{a_i} \ket{\Phi_i} \bra{a_i} \bra{\Phi_i}
 \label{densitymatrixEq2}
\end{equation}
i.e., into a statistical mixture of the various outcomes where only one result is realized in any individual experiment with probability  $|c_i|^2$.

\bigskip

Let's consider a practical example of such a measurement of the spin of a particle using a measuring device called a Stern-Gerlach apparatus. To measure the $S_z$ component of a spin-1/2 particle, we pass a beam of particles through a magnetic field that gives the particle a kick in the direction perpendicular to the path of the beam.  The measurement is obtained by noting the vertical position of the spot on the screen.  In the ideal case, only 2 spots can be seen, always aligned with the direction of the magnet.

\bigskip

Let's be a bit more precise about these statements.  For this measurement, the Stern-Gerlach set-up  will have a magnetic field along the $z$ axis. For a ``hard" measurement, $B_z$ will need to have a significant gradient along the same $z$ axis $B'= dB_z/dz$. Let the initial wave function of the incoming spin-1/2 particles with magnetic moment $\mu$ be:
\begin{equation}
\ket{S_{x}=1/2}=\frac{1}{\sqrt{2}} \lbrace\ket{S_{z}=1/2}+\ket{S_{z}=-1/2}\rbrace
\label{initialwavefunctionEq3}
\end{equation}
The interaction $H=\vec{\mu} \cdot B \sim B'z \sigma_z$ couples $\sigma_z$, the quantum spin operator, with the $z$ coordinate of the atom carrying the spin and $exp(iHt)$ (we use $\hbar=1$)  shifts the momentum $P_z$ of the atom serving here as a ``pointer." The initial atomic beam moving along the $x$ axis, say, splits into an upward moving and downward moving beams with $S_z =1/2$ and 
$S_z =-1/2$, respectively. In other words, application of the Schr\"odinger equation for the Stern-Gerlach
measurement produces the pure wavefunction: 
\begin{equation}
{1 \over
2}\lbrace\ket{S_z=1/2}\ket{SPOT_{up}} + \ket{S_z=-1/2}\ket{SPOT_{down}}\rbrace
\end{equation}
 for the combined particle
and measuring device which is just the spot on the photographic plate.  However, for an individual particle, we always see it only in the upper spot or in the lower spot, i.e. the atoms in the two beams eventually manifest via``counts" in the up and down portions of the apparatus. 
To describe this mathematically, von Neumann taught us to apply the reduction postulate.  That is, the wavefunction should be mixed, either 
\begin{equation}
|S_z=1/2\rangle|SPOT_{up}\rangle
\end{equation}
\centerline{\bf \it OR}
\begin{equation}
|S_z=-1/2\rangle|SPOT_{down}\rangle
\end{equation}
\noindent depending on the result of the measurement.  
However, before the irreversible recording but after the splitting of the beam into the downward and upward components we could, in principle, reverse the process: refocusing the two beams into one and by interference reproduce the initial $\ket{S_x=+1/2}$ state. (A close analog of this is routinely done with polarized photons.)

\bigskip

In summary, the outcomes have the following properties: 
\begin{itemize}
\item It can only be a particular value, called the eigenvalue $a_i$.
\item A particular outcome $a_i$ appears at random, with probability depending only on the initial state of the measured system and is independent of the details of the measurement. 
\item The measurement leads to the (true or effective, depending on one's preferred interpretation) collapse of the wave function of the measured system onto another state, called the eigenstate $\vert a_i\rangle $. 
\end{itemize}

\bigskip

One naturally asks: ``At what point does one stop describing the
chain of measuring devices quantum mechanically and describe it
classically?"  Nothing tells us where to  place the cut between these two incompatible descriptions.   This is the unsolved quantum measurement paradox.

\bigskip



Nevertheless, we know  that at {\em some} stage of the measurement process, an ``imprint" is formed, say a grain in the photographic plate exposed to the atoms as a result of some ionization process. At this point, the upward and downward branches ``decohere" and therefore can no longer interfere.
The above description of the ``collapse" does not explain how (or possibly even why) one  particular branch is ``picked out," becoming eventually the  observed unique result of the experiment with all other results and corresponding branches getting ``cut off."  Quantum mechanics and the linear Schr\"odinger equation cannot address these questions.

\bigskip

  Over the years, many ``explanations" of the collapse have been suggested, such as a stochastic interaction localized in configuration space, the GRW 
mechanism and its generalizations by Pearle and others \cite{pearle}. In the ``many worlds" interpretation, all possible results of all 
experiments are realized but in different worlds. In this approach, a vast number of worlds are thereby generated by every experiment planned or naturally occurring. At every 
bifurcation point that the two worlds differ, by definition, only by the position of {\em one} specific pointer. Suppose subsequent experiments are performed at fixed 
time intervals $T$. Then the number of universes will grow as $\sim 2^{2^2}$ (operation repeated $\frac{t}{T}$ times). This growth rate seems to be unacceptably high. Already, 
after just seven steps, we have $2^{10^{80}}$ universes. Assuming a momentum space cutoff at $m(Planck) \approx{10^{19}}GeV$, the number of states in the Hilbert 
space of the entire observable Hubble size universe will be exceeded in just eight steps!

\bigskip

In the following, we reconsider the issue of the collapse in an indirect way. We suggest that basic physics prevents even ``gedanken" experiments where 
distinct ``classical" pointer states interfere.  We also elaborate on a remarkable reverse, i.e. a classical-to-quantum, evolution. We show that almost any initial semiclassical WKB state which is subjected to multiple reflections  quickly spreads out into an extended delocalized quantum state and continuous repeated collapses are needed to maintain classicality. 

\bigskip

This conclusion, namely that to avoid complete quantum ``smearing," we have to keep the system open and constantly interacting with an external ``environment," is of key importance. Simple environments that we will briefly discuss seem to be unable to do the job.  We show that some minimal richness and complexity are required. A bold and far reaching speculation  (related to some earlier work of one of us,  A.C.) is that the classical limit can be obtained {\em only} if the quantum system in question is brought into contact with an environment which is a heat bath.

\bigskip

 As for the specific mechanism, if any, which picks out the ``correct" branch in each experiment, we only mention the suggestion \cite{AGrus}~\cite{abl}~\cite{PT}~\cite{townes}~\cite{danny}
that it can be provided in the two-time formulation by specifying along with the overall initial state of the whole universe another, final, ``Destiny" state.  

\bigskip

The plan of the paper is as follows:

\bigskip

 In section \ref{sectionII}, we address the impossibility of achieving exact time reversal in a system with many degrees of freedom. This is a particular example of the difficult task of ``aiming" an initial classical state so as to become a specific final state.

\bigskip

 In section \ref{sectionIII}, we comment on the classical-to-quantum transition in any non-separable closed system of $n \geq 2$ degrees of freedom. Even if the system is initially in a well defined WKB, semi-classical state, quantum evolution and, in particular, multiple reflections at classical turning points make it completely quantum mechanical with each particle smeared almost uniformly over all the configuration space. The argument, which is presented in the context of $n$ hard discs, is quite general. Finally, we  briefly address more complex quantum systems with many degrees of freedom  and ask when can they provide an appropriate environment to the above simpler systems so that quantum spreading is avoided by continuously leaving ``imprints" in the environment.  We also discuss the possible connections with the pointer systems that are needed in the quantum-to-classical ``collapse" transitions.

 \section{Time Reversal Transformation and Precise Guiding in General}
\label{sectionII}
The time-reversal  transformation $r_i(t) \rightarrow r_i(t)$ and $p_i(t) \rightarrow -p_i(t)$ is anti-canonical, reversing the sign of the $(p_i,r_i)$
Poisson bracket. The corresponding quantum mechanical transformation is anti-unitary and cannot be implemented by any Hamiltonian evolution
$U=\exp{(-iHt)}$ over all the Hilbert space.
It is instructive to see how the following attempt to realize a time-reversal actually fails. Consider a system of spin-less charged particles, all with the same charge to mass ratio: $e_i/{m_i} =e/m=const$, and moving in the $x, y$ plane. Let us  turn on, for a short time $\delta(t)$, a strong transverse magnetic field $B_z$  such that $w_{Larmor} = eB_z/(mc) = \frac{\pi}{\delta(t)}$. This would appear to achieve time reversal. It reverses all velocities by making each particle describe a semi circle of radius $a_i=v_i/(w_{Larmor})$.  As $B_z\rightarrow \infty$ the gyration radiuses tend to zero and the coordinates remain fixed as they should under a time-reversal transformation. However, the impulsive turning on and off of $B_z$ induces huge electric fields. Thus, let the system be inside a large circle of radius  $R >|r_i|$.  In order to reverse the momenta of all particles at any location $r_i$, let $B_z$ be turned on over the whole area. Faraday's induction generates at a  distance $r$ from the center, azimuthal $E$ fields, $E_\phi \sim r dB_z/(dt)$,  which are much larger than $B$. While the induced E reverses sign between the turning on and turning off phase, the particles momenta rotate in the interim so that the coordinate shift due to the E field, which increases with $r$, will not cancel and spoil the required time reversal transformation.

\bigskip

In a quantum field theoretic framework, this difficulty reflects the fact that the transverse vector potential $A$ which yields $B$ as its curl, and 
$E \sim dA/(dt)$ are canonically conjugate. The impossibility of reversing a coordinate without reversing also the corresponding canonically conjugate momentum manifests here for both the charged particles and the electro-magnetic fields.

\bigskip

If we focus only on the spins of the spin-1/2 ``electrons," an appropriate $B_z$ field will indeed reverse $\sigma(x)$ and $\sigma(y)$ (that is, the spin 
components in the plane).  It is the same as the above which reverses the ordinary momenta/ angular momenta, if the gyromagnetic ratio would have been $g=2$ 
exactly. It will then fail to reverse the commutator $[\sigma(x), \sigma(y)]=i\sigma(z)$ which is clearly another relevant operator. 

\bigskip

 Can we still achieve time reversal for {\em one} specific state? The following suggests that even this is, in general, impossible for systems with a large number of degrees of freedom. Classically, this is related to the second law of thermodynamics, but it is also relevant to question of the collapse. Specifically, if we apply this putative time-reversal transformation in the context of the Stern-Gerlach set-up to one particular state at the time where a grain has already developed on the photographic plate in the, say, upper section of the instrument, then it should allow us to ``undo" the collapse of this state.  We show that this task is impossible. 

\bigskip

We next present the concrete model in the framework of which most our discussions will be made. It is a closed system of $N$ hard spheres or discs, of radius $a$ and average separation $l$, located inside a large reflecting  sphere of radius $R=N^{1/3}l$ or circle with $R=N^{1/2}l$ in the 3 or 2 dimensional cases respectively. The mean free path for collisions $(n_t.\sigma)^{-1}$ then is:
\begin{equation}
 l_{mfp}=1/(n_ta^2)=l^3/(a^2)\; {\rm for} \; d=3, \; {\rm or} l^2/a \; {\rm for} \;  d=2
\label{meanfreepathEq4}
\end{equation}
where the number density of targets $n_t$ is $\sim l^{-3}$ and $l^{-2}$ in the $3-$ and $2-d$ cases.

\bigskip

The basic question posed is: 

\begin{quote}
``Can we fix at time t=0 the locations and velocities $r_i(0)$ and $v_i(0)$ with sufficient accuracy so that after time
$T$, all the $N$ coordinates $r_i(T)$ are within small errors $\epsilon_i \rightarrow 0$ at pre-destined `goals' $r_i^g$?"
\end{quote}

\bigskip

The following question then arises: suppose we try to interfere the two branches of the Stern Gerlach experiment after the spin carrying particle caused a grain to develop. We then need (at least) to translate via some evolution Hamiltonian over a time $T$,  the section of the upper half of the apparatus around the putative grain to the symmetric down part of the apparatus with its putative grain. Furthermore, each molecule, atom etc. in one branch should precisely overlap the corresponding molecule, atom etc in the other. Otherwise, $\Phi_{down} (Q)$, the down pointer and $\exp\lbrace{iHT}\rbrace \Phi_{up} (Q)$ the translated up pointer, become orthogonal and will not interfere.

\bigskip

The complexity of the grain embedded in the up/down parts of the apparatus prevents us from meaningfully addressing this problem. Hence we
use the simpler model system of $N$ hard balls/discs. The initial state of the $N$ objects with centers at $(r_i(0)); i=1,2...N$ represents the pointer state
 $\ket{P_u}$ of the ``up branch" in our idealized Stern-Gerlach type experiment. Likewise, a different set of ball locations $r_i^g; i=1,2...N$ clustered at 
some distance $L$ away from the first set, represents the pointer state of the down branch, $\ket{P_d}$. In order to obtain optimal interference between this $N$ ball state $\ket{P_d}$ and the 
state $\ket{P^{'}_{u}}=  \exp(iHT)\ket{P_u}$  obtained by shifting the original up grain state, we need to have the following condition: $\bra{P^{'}_{u}}P_d\rangle= 1$.
 Neglecting correlations, this is the product of overlaps of the $N$ parts:
\begin{eqnarray}
 \bra{P'_u}P_d\rangle &\sim& \bra{1}1'\rangle\bra{2}2'\rangle...\bra{N}N'\rangle \sim\nonumber\\
&\sim& (1-\delta)^N = exp(-N\delta)
\label{NpartoverlapsEq.5}
\end{eqnarray}
Hence we need that each of the final $r_i(T)$ will be close to the desired ``goal" location $r_i^g$ to within a fraction $\delta =1/N$ of the average distance between the balls. The center of each ball is represented by a wave packet of effective size which can be as small as $\lambda=h/(mv)$. The condition for interference then becomes:
\begin{equation}
 |r_i(T)-r_i^g| \leq \delta\lambda \leq \lambda/N
\label{intereference6}
\end{equation}
where $\lambda$ can be many orders of magnitude smaller than the classical counterpart which is the actual experimental uncertainty of the 
locations of the balls. 

\bigskip

It is amusing to note that the last condition is invariant under the following ``scaling" operation. Instead of $N$ balls (discs), consider
$N'=N/k$ ``super-balls/discs" each consisting of $k$ of the original balls/discs.  The $k$ fold increase of the masses and corresponding decrease of
the de-Broglie wavelength matches the decrease in  the number of balls/ discs - yielding the same condition.

\bigskip

Such a precise guiding of the $N$ particles by fine-tuning the velocities is impossible in either the semi-classical case (discussed in this section) or in the fully quantum cases addressed in the next section. This impossibility traces back to the classically chaotic behavior of generic systems and our $2-d$ billiards in
particular. Thus, let us assume that an ``error", i.e. a deviation from the ``correct" classical path occurred for {\em one} of the discs $D^i$. After suffering $(n-1)$ collisions, it should have entered the n$^{th}$ collision with impact parameter $b^{0(i)}_n$, relative to the center of the other disc, but
 $b^i_n=b^{0(i)} \pm \delta b^i_n$. This deviation involves errors in the locations of the centers of both discs. In a simplified, conservative
estimate of the accumulating errors in successive collisions, we ``freeze" N-1 discs allowing only one active disc $D^i=D$ to bounce between them. Similar errors obtain when all $N$ discs are moving.

\bigskip

Tracing the path of the``active" disc $D$, we see that the shift, $\delta b_n$, causes a change in the direction of $D$ as it recoils after the collision by:
\begin{equation}
 \delta \theta_n \sim  (\delta b_n)/a
\label{discDrecoilEq7}
\end{equation}

The disc then travels a distance of $|r_{n+1}-r_n| \sim l_{mfp}$ before encountering the pre-designated partner in the next (n+1)$^{th}$ collision.
 Hence, at the time of this collision, the shift, or error, in the impact parameter will become:
\begin{equation}
 \delta b_{n+1} \sim  \delta \theta_n |r_{n+1}-r_n| \sim \delta b_n \, l_{mfp}/a=(l/a)^2 \cdot \delta b_n
\label{impactparamshiftEq.8}
\end{equation}
where we used the m.f.p of eq.$ \ref{meanfreepathEq4}$ for collision in $2-d$. Thus the errors keep amplifying by an enhancement factor $c= (l/a)^2$  at each collision (and by an enhancement factor of $c= (l/a)^3$ in the three dimensional case).

\bigskip

The exponential increase with $n$ of $\delta b_n$ with a positive (and large!) Liapunov exponent $(2-3) log(l/a)$ means that our system and most  multi-component classical systems are ``chaotic": the path of any particle, and of the full system, is sensitive to tiny changes in the initial conditions. 
This is closely related to the issue of irreversibility and the second law of thermodynamics.

A famous example involves a ``gas" of, say, $2-d$ hard discs, initially confined to some part of a given domain (e.g. a billiard table). By removing a partition, it is allowed to ``expand," and, in a time $T$, it fill up the whole area.  To implement a time reversal transformation, we need to reverse at time $T$ all the velocities so that the system retraces  its evolution back via multiple collisions and all discs reconvene at time $2T$ to the original smaller region. It has been suggested \cite{AharonovEnglertOrloff},  that if one were to fail to reverse the velocity of even just {\em one} among the many $N$ discs, then this would suffice to destroy the specific pattern of collisions required to implement a time reversal transformation.

\bigskip

The case of interest here in which we have small errors, $\delta r^i(0)$ and $\delta v^i(0)$, in the initial locations and velocities of all the discs
is more readily analyzed. As seen above, these errors are exponentially amplified proportional to $c^n$ with $c \gg1$ and $n$ the average number of 
collisions suffered by any disc among the $N$ discs in our system. Once the uncertainty in the impact parameter exceeds $2a$, with $a$ the radius of the discs, 
the discs will start missing altogether the `partners' that they were ``designated" to collide with.  As a consequence, the evolution will drastically change. Hence, there will be no `re-gathering' of all the discs or the molecules of an ideal gas  in the smaller region.

\bigskip

For the case of colliding hard discs, these difficulties are very concrete. Let us denote by $n_{crit}$ the number of collisions that
need to happen before the``missing of partners" and therefore the setting in of  complete chaos. We will argue that $k$, the actual number of collisions
expected during the expansion into the larger domain (or in the time reversed case  during the retraction to the smaller domain)
does exceed $n_{crit}$ thereby prempting the putative time-reversal and violation of the second law. 

\bigskip

Let $v$ be the average rms velocity of the discs/molecules. The time $T$ required for the gas to expand into the larger enclosure is then at least $R/v$
and
\begin{equation}
k \sim \frac{T v}{l_{mfp}} \,{\scriptstyle{>\atop\sim}} \,\frac{R}{l_{mfp}}.
\end{equation}
 Collisions will then be suffered during this time by any molecule/disc.

\bigskip

 After this minimal time, the motion of the gas is still quite ordered and the relevant actual equilibration time, $T$, is in fact much longer. Thus, let the initial and final enclosures be concentric circles of radia $R/2$ and $R$ respectively. After the inner ring is removed, spherical ( circular) pressure waves propagate outwards with the  velocity of sound, $v_{sound} \sim v/(2^{1/2})$, arriving at $R$ after a time $\sim{T}$. Since the mean free path for collisions is much shorter than $R$, then the molecules in concentric shells within the small circle map into corresponding but doubled in size concentric shells in the larger enclosure with the radial ordering of the different shells preserved.

\bigskip

To get a truly mixed states manifesting classical irreversibility more strongly, we need a much longer time and therefore many more collisions will ensue. If we ask that each individual disc diffuse through the system by a random walk, then we have $k=(R/(l_{mfp}))^2$. We expect the actual $k$ to
be intermediate between this and the linear $k=R/l_{mfp}$ lower bound value. For fixed number density of discs $\sim l^{-2}$, $R$
increases as $N^{1/2}$ and hence, $k$ scales as N$^{1/2}$ or as $N$ respectively.

\bigskip

To avoid ``missing partners" after $k$ collisions, we need that $c^k\delta b_0 < a$. Using the enhancement factor $c=(l/a)^2$, this then implies that 
$(\frac{l}{a})^{-2k}  \geq{\delta b_0} $. With $k$ and $(\frac{l}{a})^2$ of order a hundred, this requires incredible precision and therefore small $\delta b_0 \leq{10^{-200}}a$. This is classically impractical and quantum mechanically outright impossible. Such fine tunings of the positions of all discs requires, by the uncertainty principle,
huge momentums and energies, which in the presence of gravity, will cause the system to {\em collapse into a black hole}.


\section{The classical to quantum diffusion of classically chaotic systems}
\label{sectionIII}

We have argued above that quantum uncertainty thwarts efforts to achieve time reversal in any system with a large number of degrees of freedom. Yet the argument we presented in \S\ref{sectionII} was incomplete and potentially inconsistent. It treated the time evolution of the $N$ discs in our model system {\em classically} and quantum mechanics manifested only via our inability to fix the initial conditions with the accuracy required. We address this in this section and show that the situation is no better using a completely quantum mechanical treatment. We do this by studying the quantum mechanical evolution of the system of $N$ hard discs with (the center) of each disc starting out in narrow Gaussian WKB wave packet. We find that closed systems which are classically chaotic, completely lose their classical nature, after a moderate number of ``periods" 
(or average number of collisons of each disc).  Consequently,  the center of each disc spreads out over the whole volume of the system.

\bigskip

 We show that the wave function of a particle (representing the center of our mobile disc $D$ ) spreads upon reflection from the other discs  according to classical expectations. Specifically, we can follow two extreme rays at the edges of the wave packet enclosing most of the norm of the impinging wave. 
These two rays then reflect via Snell's law and continue to describe the edges of the system at later times. This leads to an exponential increase of the
transverse, x space, width of the wave function  which we denote by $\Delta$. We require that our initial wave packets satisfy:
\begin{equation}
 \lambda << \Delta(t=0) << a
\label{initwavepktsrequirementEq9}
\end{equation}
and we demand that $\Delta(t=0 >> \lambda \sim 1/p$ with $p=Mv$ the momentum in order that the uncertainty in the transverse momentum be far smaller than the momentum of the disc. Finally, we also require that the ordinary diffusive broadening of the Gaussian packet during the travel between subsequent collisions be small compared with $\Delta$, leading to:
\begin{equation}
  \delta \Delta^2 \sim t/M \sim l_{mfp}/p_0 << \Delta^2
\label{diffusivebroadeningEq10}
\end{equation}
with $p_0$ the central momentum of the wave packet in momentum space of width:
\begin{equation}
  \Delta p \sim 1/\Delta_0
\label{DeltapEq11}
\end{equation}

If after $n$ collisions, the transverse (configuration space) width of the wave packet grows to $\Delta_n \sim a$, then we cannot be sure if the disc will indeed collide in the $n+1$th' step with the ``designated disc" on the classical trajectory. Rather, we will only be able to assign probability amplitudes for this or any other collision to occur. The wave function becomes a complicated path integral over all histories of (potentially correlated) collisions. Any prospect of controlling or guiding such a system is clearly lost at this point.

\bigskip

As in the previous section, we first consider one point mass moving in a background of $N-1$ stationary discs (the effective hard core radius is now $2a$, but still referred to as $a$). The initial wave packet is a Gaussian of width $\Delta_0$ in both the $x$ and y direction. It moves along the $x$ axis and eventually collides with the first disc (which for convenience we center at the origin) with an impact parameter $b \pm \Delta_0$. The corresponding angular location of the collision on the rim of this disc is therefore:
\begin{equation}
 \phi_0 = arcsin(b/a)
 \label{collisionangularlocationEq12}
\end{equation}
and angular width
\begin{equation}
 \Delta\phi_0 \sim  \Delta_0/a.
\label{angularwidthEq13}
\end{equation}
Since we assumed that $\Delta_0 << a$ and $a << l$ with l the average disc separation, the collisions with any given disc are independent and can be
 treated exactly. Using the partial wave ,exp($im\phi)$ Fourier decomposition approriate for the $2-d$ geometry, the exact phase shifts upon scattering 
from the hard disc are found to be:
\begin{equation}
 \delta_m(p_0) = arctan(J_m(p_0a)/N_m(p_0a))
\label{exactphaseshiftsEq14}
\end{equation}

 The asymptotic behavior of the above Bessel (and Neumann) functions in the relevant limit of jointly large index and argument implies that the 
incoming wave packet is reflected according to Snell's law. The nontrivial aspect 
 which underlies the subsequent
discussion, is that also the incident ``limiting rays," impacting at $b_0 \pm \Delta_0$, also reflect according to Snell's law.

\bigskip

This is indeed expected since our packet consists of many waves and we could apply the above snell reflection to each one separately. Stating it differently,  we can locally use  in each region the Huygens construction to build the reflected waves.  It then follows that the initial square Gaussian packet becomes upon
reflection a wave packet in $r$, the radial coordinate measured relative to the center of the reflecting disc (the origin by assumption here) of width 
$\Delta r = \Delta_0$, and in the angle  $\phi$ with width $\Delta\phi$ given by eq. (\ref{collisionangularlocationEq12}) above.
Simple geometry then implies that upon further propagation, this wedge-like wave packet, maintains
its angular opening $\Delta\phi$ and broadens in the transverse direction. Thus, by the time our wave packet collides with the next disc, at an average
distance $r \sim l_{mfp}$ away, the transverse width and corresponding impact parameter and angular uncertainties will be amplified just like the errors in the classical case discussed at length in the previous section:
\begin{equation}
 \Delta_1 \sim  \Delta b_1 \sim c \cdot \Delta_0
\label{Delta1amplificationEq15}
\end{equation}
\begin{equation}
 \Delta\phi_1 \sim  \Delta b_1/a \sim c \cdot \Delta \phi_0
\label{DeltaBetaamplificationEq16}
\end{equation}
with $b_1$ and $\phi_1$ measured relative to the second disc and the enhancement factor $c$ given by:
\begin{equation}
 c= l_{mfp}/a \gg 1
\label{enhancementEq17}
\end{equation}

We can repeat the above getting further spatial and angular broadening so that after $n$ collisions:
\begin{equation}
 \Delta_n \sim c^n \Delta_0.
\label{broadeningEq18}
\end{equation}

After a relatively small number of collisions
\begin{equation}
  n \sim  ln (a/(\Delta_0)/(ln \, c)
\label{collisionsEq19}
\end{equation}
We find $\Delta_n \sim a$ indicating the onset of genuinely quantum mechanical evolution and loss of the initial WKB wave packets.

\bigskip

 One may worry  that the WKB approximation is inappropriate for the case of infinitely hard discs as the ``h" expansion around classical paths utilizing the Wigner function and Moyal brackets involves derivatives of the potential and such systematic expansions break down. However, we have solved the scattering problem exactly and do not need such expansions here. 

\bigskip

Our concrete result and more general arguments suggest that the catastrophic spreading of the WKB wave function is a general and universal feature. 
The amplification factor evaluated more carefully, $2r_0\gamma/(a^2-b^2)^{1/2}$, can be written in the following suggestive form:
\begin{equation}
 c=\frac{r_0\gamma h}{p_0} \, 
\Delta^2 \delta(p,b)/\,
\Delta b^2 \sim  \frac{r_0 \gamma}{p_0}2 \,
\Delta^2 \; S_c(p_0,b)\,
\label{evalamplificationfactorEq20}
\end{equation}
with $S_c(p_0,b)$ the classical action of the trajectory with ( central) momentum $p_0$ and impact parameter $b$ and $\delta(p,b)$ the corresponding phaseshift for angular momentum $n=p_0b$.

\bigskip

We note that the broadening of the wave packet in directions perpendicular to that of the reflection is expected. In general, the WKB approximation is vulnerable around the (classical) turning points where $p_x$, say, $\rightarrow{0}$ and the particle (which now is also a wave of very large wave length!) ``hovers" around there for a long time. If the motion is one dimensional, the WKB wave function simply accumulates an extra $\pi/4$ phase due to the Airy function  interpolation~\cite{landau}. However transverse motion can be strongly affected.

\bigskip

In the general, multi-disc situation, the particular disc $D$ that we focused on does not move all by itself. Each time it reflects from some other disc $D_i$, it leaves an ``imprint" in the other recoiling disc. All these entanglements are however artifacts of separating the motion of one or a few of the
discs and cannot save the closed system from the catastrophic quantum spreading. We should also consider the wave function $\Psi(r_1,r_2,..r_N;t)$
representing all the $N$ discs and which depends on the $2N (x_i,y_i)\,\, i=(1,..,N)$ coordinates.  To start with, at $t=0, \Psi$  factorizes into $N$ well defined WKB wave packets. The time evolution with many collisions does not only make $\Psi(r_1,r_2..r_N;T)$ with $T >> R/v$ more complex. Rather the probability of finding any disc at a location $r_i$ is completely smeared out and no longer concentrated around the classical path on the $N$ particle system.

\bigskip

 It should be emphasized that despite the superficial similarity, the quantum smearing is very different from the chaotic behavior of the corresponding classical $N$-disc system. They are similar in that in both cases we do not know where any one of the discs is located after some time $T$ (the time over which each disc suffered the required critical number of reflections). However, while classically it is ``just" an issue of our inability to carry-out calculations with infinite precision, in the quantum system, we cannot know the locations even in principle, regardless of the computational resources available.

\bigskip

It is instructive to point out a gedanken experiment which can distinguish the two situations. Let us  probe the system at time $T$ by sending in an intense lazer beam in the transverse $z$-direction  (with a wavelengt which is larger than $l$, the average separation between the discs, and smaller than $R$ the size of the entire system). Also, the duration of the pulse is shorter than the relatively long time $l/v$ which it takes the slow Discs to travel between successive collisions. The resulting diffraction picture in the quantum case   will correspond to a homogeneously distributed matter whereas in the classical case, it will reveal the instantaneous locations of all the $N$ discs, albeit in a blurred fashion.

\bigskip

Since all the above aspects are generic, we do not believe that the spreading can be avoided by any further additional reflectors or potentials and/or even the applications of a time dependent potential.

\bigskip

This quick destruction of semi-classical behavior of any closed system with initial quasi-classical WKB states is more then another obstacle impeding the goal of undoing the collapse. We know that classical systems, chaotic or not, do remain classical. We therefore must have some mechanisms for ensuring this, which in effect may necessarilly involve many ``mini-collapses!" 

\bigskip

 It is even more amusing to note that the above mentioned quantum smearing does not happen in systems with one degree of freedom. Likewise it should not happen when, by an appropriate canonical transformation the system can be separated into $N+N$ action-angle variables. Such exactly solveable systems would maintain the original well-defined narrow WKB form. Only very few systems, such as those with harmonic interactions, are of this kind. We note that certain statistical mechanics and field theoretic systems (with infinitely many degrees of freedom) are solvable though generally only in $1+1$ dimensions. Could it be that at some fundamental level, when formulated in terms of some new magical degrees of freedom, most systems not only have a simple Lagrangian like that of QCD, say, but are also exactly solvable?  If this was the only way out of our quantum spreading dilemma, then searches for exactly solvable models would gain motivation far beyond what they enjoy at present. We note that, most likely, quantum broadening can be avoided in more mundane ways.

\section{Summary and Conclusions}
\label{sectionIV}
 Aspects of the material discussed above is not really new and is ``well known to those who know it well." Indeed wave function broadening and complete quantum smearing after a short period of time of chaotic {\em astronomical} systems, such as Hyperion, Saturn's asymmetric moon, has been claimed to arise  in numerical simmulations, if the astronomcal system was completely closed. Still, having the phenomenon illustrated in the present simpler context may be useful. \footnote{though simplicity, like beauty, is often in the eye of the beholder}
 How is this evaded in the real world? The standard (and correct answer)  results from the fact that actual systems are open and their evolution often leaves marks in the environment.  This thereby stops the quantum spreading. Still it would be nice to have concrete examples as to how this is being achieved.

\bigskip

  In trying to address this, we have studied the following ``natural" environment: let us replace our $2-d$ hard discs of size  $a$ by an infinitely thin yet equal mass ring of the same radius. Rather than having just the $2N (x_i;y_i)$ degrees of freedom associated with the $N$ centers of the rings, we now also have the $N$- fold tensor products of the Hilbert spaces of $N$ closed strings, with each one associated with one of the $N$ rings. Whenever two rings collide, the event is ``recorded" via the elastic phonons emitted from the point of contact at the time of the collision. We suspect however that notwithstanding the extreme richness of our system which even for one Disc is that of a closed string with its infinitely many modes, this will not evade the eventual quantum broadening. The point is that we do not really leave an irreversible mark as in the dissipative case. The elastic wave may converge again at some point and kick the colliding rings back so that the information is not erased and we simply have just a much bigger and more complex system. 
It may be that the key is to have a very dense set of energy levels as in generic macroscopic systems so that the times for recurrences of the above types are infinitely long. It also may be that we need a genuine thermodynamic heat bath and dissipation. Clearly further research is needed here.  

  \bigskip

\noindent {\bf Acknowledgments:} We thank Yakir Aharonov and Federico Faggin for discussions.

\end{document}